\documentclass[aps,prd,reprint,a4paper]{revtex4-1}
\usepackage{graphicx}
\begin{document}
\title{Sequential recombination
algorithm for jet clustering and background subtraction}
\author{Jeff Tseng}
\email{j.tseng1@physics.ox.ac.uk}
\author{Hannah Evans}
\affiliation{University of Oxford, Subdepartment of Particle Physics,\\
Denys Wilkinson Building, Keble Road, Oxford OX1 3RH, United Kingdom}
\date{\today}

\begin{abstract}
We investigate a new sequential recombination algorithm which effectively
subtracts background as it reconstructs the jet.
We examine the new algorithm's behavior in light of existing algorithms,
and we find that in Monte Carlo comparisons,
the new algorithm's robustness against collision backgrounds
is comparable to that of other jet algorithms when the latter
have been augmented by further background subtraction techniques.
\end{abstract}

\maketitle


Collimated jets of particles are a distinctive feature of high energy
elementary particle collisions and are often taken to indicate the presence
of ejected quarks or gluons, particles normally shrouded by the effects
of quantum chromodynamics (QCD).
Jet reconstruction therefore plays a prominent
role in event analysis, and as the search for new physics
breaches new thresholds in energy and jet multiplicity,
understanding jet reconstruction itself has taken on new
importance.  This importance is especially true in the study of
highly relativistic (``boosted'') objects, in which evidence of
heavy or exotic particle production and decay can be discerned in a
jet's substructure.  Experimental results on jet substructure
have been published by the CDF~\cite{Aaltonen:2011pg},
ATLAS~\cite{ATLAS:2012am,Aad:2013gja}, and
CMS~\cite{Chatrchyan:2013rla} experiments.

A standard class of methods for jet reconstruction in hadron collider
experiments is the sequential recombination algorithm.
Different varieties of
this algorithm usually are rooted in physical or geometric
considerations, such as QCD splitting functions for the $k_T$
algorithm~\cite{Catani:1993hr,Ellis:1993tq},
angular ordering for Cambridge-Aachen~\cite{Wobisch:1998wt},
and collimated jet
cores for anti-$k_T$~\cite{Cacciari:2008gp}.
In this article, we consider a modified sequential
recombination algorithm which has some features seen in
the classical theory of radiation by moving charges.  The algorithm
simultaneously removes background radiation, including initial
state radiation, particles of the underlying event,
and those from the unassociated
collisions (``pileup'') which are an important feature of modern
high-luminosity colliders such as the Large Hadron Collider.

The first section of this article describes the modified
algorithm and compares it with others.
In Section~\ref{sec:perf}, we test the new algorithm on
simulated high-energy $W$ bosons with and without the presence of
pileup, and compare the results with those of other clustering
algorithms.  Comparisons are also made with further background
removal (``grooming'') techniques.

\section{Algorithm}
\label{sec:alg}

In a typical sequential recombination algorithm, we start with a
set of 4-vectors (``clusters'') which could represent the momenta of particles,
calorimeter energy deposits, or previously clustered 4-vectors.
The initial clusters are assumed to be massless.  
Interjet distances
\begin{equation}
d_{ij} = \min[p_{Ti}^r,p_{Tj}^r]\left(\frac{\Delta R_{ij}}{R}\right)^2
\end{equation}
are calculated for each pair of clusters $i$ and $j$,
as well as beam-jet distances
\begin{equation}
d_{iB} = p_{Ti}^r
\end{equation}
for each cluster $i$.  In these formulae,
$p_{Ti}$ is the momentum of cluster $i$ transverse to the beam, and
$\Delta R_{ij} = \sqrt{(\Delta y_{ij})^2+(\Delta\phi_{ij})^2}$, where
$\Delta y_{ij}$ is the rapidity difference and $\Delta\phi_{ij}$ the
difference in azimuthal angle.  $R$ is a jet scale parameter which
defines the maximum $\Delta R_{ij}$ for clustering pairs.
The parameter $r$ is 2 for the $k_T$ algorithm, 0 for
Cambridge-Aachen, and $-2$ for the anti-$k_T$ algorithm.
If the smallest
distance is a $d_{ij}$, the pair is merged, often by adding the 4-momenta.
If the smallest distance is a $d_{iB}$, the cluster
is deemed an independent jet and removed from further consideration.
These steps are repeated until all clusters have been deemed jets.

We now consider a new algorithm with distance measures
\begin{eqnarray}
d_{ij} & = & \frac{1}{4}(m_{Ti}+m_{Tj})^2\left(\frac{\Delta R_{ij}}{R}\right)^3,
\label{eq:scdij} \\
d_{iB} & = & m_{Ti}^2,
\end{eqnarray}
where $m_{Ti}=\sqrt{m_i^2+p_{Ti}^2}$
is the ``transverse mass'' of cluster $i$, and $m_i$ is its
mass.\footnote{An earlier version of the algorithm used
$E_T$, the transverse energy, instead of $m_T$, which has the
advantage of being longitudinally invariant.  The differences are small
for the tests presented in this article.}
The coefficient 1/4 has the effect that
$d_{iB}<d_{ij}$ whenever two jets with the same $m_T$
are separated by $\Delta R_{ij}>R$.  If $m_{Ti}<m_{Tj}$, we also have
$d_{iB}<d_{ij}$ whenever $\Delta R_{ij}>R$,
and therefore $R$ is, as in other algorithms, the
maximum $\Delta R_{ij}$ between clusters that can be merged.
The algorithm is collinear and infrared-safe by construction.
We call this algorithm ``semi-classical'' (SC) by analogy with
the classical angular distribution of radiation from moving
charges, which depends on the relativistic
boost factor $\gamma\sim E_i+E_j \sim m_{Ti}+m_{Tj}$
(replacing energy by longitudinally invariant quantities),
and has different exponents for the energy and angular
($\Delta R_{ij}$) factors.
The cubic angular exponent in Equation~\ref{eq:scdij} arises from
considering isotropic, massless emissions in the parent body's
rest frame.
Different exponents for the angular factor have been tested,
and, for the most part, merely allow clusters to merge with
larger or smaller $\Delta R_{ij}$, thus changing the size
of the resulting jets.

\begin{figure}[tbp]
\includegraphics[width=3.4in]{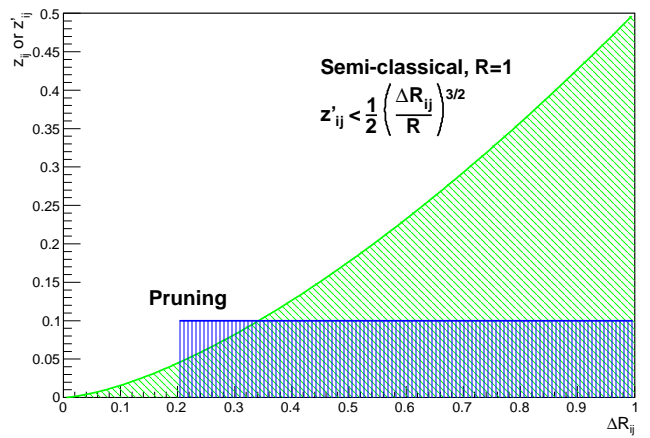}
\caption{Comparison of the semi-classical algorithm with pruning.
The diagonally hashed region indicates mergings rejected by the
semi-classical algorithm, while the vertically hashed region
is for pruning.
The pruning parameters are taken from~\cite{ATLAS:2012jla}.}
\label{fig:zr}
\end{figure}

The new energy factor, on the other hand, changes the way
in which clusters are merged and set aside as jets, when compared
with other algorithms.  The $k_T$ algorithm, for instance,
starts by merging soft clusters, as one would expect for an
algorithm which attempts to reverse the presumed history of $1\rightarrow 2$
splittings in the jet.  At the same time, the $k_T$ algorithm avoids the
perceived problem of the JADE
algorithm~\cite{Bartel:1986ua,Bethke:1988zc}, with distance measure
\begin{equation}
d_{ij}=E_iE_j(1-\cos\theta_{ij}),
\end{equation}
which can allow large angle clusterings of very soft pairs.
The semi-classical algorithm also starts by merging soft pairs,
though the raised $\Delta R_{ij}$ exponent clusters some harder
clusters sooner if they are sufficiently close.
Large angle clusterings are suppressed by the
$R$ scale and beam clustering.  
However, the most significant
difference in behavior between the semi-classical and $k_T$ algorithms is that for
sufficiently large $\Delta R_{ij}$ (though still with $\Delta R_{ij}<R$), the
comparison with $d_{iB}$ prevents a number of soft clusters from merging with
high-$m_T$ clusters when
\begin{equation}
  z'_{ij} \equiv \frac{m_{Ti}}{m_{Ti}+m_{Tj}} <
  \frac{1}{2}\left(\frac{\Delta R_{ij}}{R}\right)^{3/2}.
\end{equation}
As a result, while $R$ defines the maximum extent of a jet
in the semi-classical algorithm, it is possible for two jets to be
separated by $\Delta{R}_{ij}<R$.  Moreover,
the actual jets are likely to be narrower than $R$,
with higher $m_T$ associated with narrower jets.
This behavior is similar to that of jet ``pruning'',
which vetoes mergings which satisfy the two conditions
\begin{eqnarray}
z_{ij} \equiv \frac{\min(p_{Ti},p_{Tj})}{|\vec{p}_{Ti}+\vec{p}_{Tj}|}
  & < & z_{cut}, \\
\Delta R_{ij} & > & D_{cut},
\end{eqnarray}
and discards the softer of the two clusters~\cite{Ellis:2009su,Ellis:2009me}.
Figure~\ref{fig:zr} compares the two methods, with pruning removing
the rectangular region in the $(\Delta R_{ij},z_{ij})$ plane, while
the semi-classical algorithm additionally removes some soft clusters
at small angles as well as harder clusters at large $\Delta R_{ij}$.  
These clusters become stand-alone jets
separated by $\Delta R_{ij}<R$.  

It is interesting to note that the area of a semi-classical jet,
according to either the passive or active area
definition of \cite{Cacciari:2008gn}, is zero.  This zero area
is another reflection of the effect of the algorithm's ``pruning'' of
background; from this perspective,
further area-based background subtraction is
redundant.  The effect of this pruning will be
evident in the next Section's tests with simulated events.

\section{Monte Carlo tests}
\label{sec:perf}

Initial studies of the semi-classical algorithm with boosted objects
have been performed using the {\sc Pythia}
(version 8.170) Monte Carlo
generator~\cite{Sjostrand:2006za,Sjostrand:2007gs}.  
Single hadronically decaying $W$+parton events were generated with $W$
$p_T >500\;{\rm GeV}/c$ at $\sqrt{s}=8\;{\rm TeV}$.
Non-neutrino particles were then
collected into $0.1\times 0.1$ $\eta-\phi$ cells out to $|\eta|<5$, where
$\eta=-\ln[\tan(\theta/2)]$ is pseudorapidity.
Up to an average of 25 QCD minimum bias
events, using Tune 4Cx~\cite{Corke:2011yy} and the CTEQ6L1 parton distribution
functions~\cite{Pumplin:2002vw}, were overlaid as ``pileup'',
assuming the same interaction vertex.  
Only cells with
energy greater than $0.5\;{\rm GeV}$ were considered for jet clustering.
Jets were then found using the $k_T$,
Cambridge-Aachen, and anti-$k_T$
algorithms implemented in {\sc Fastjet} version 3.0.4~\cite{Cacciari:2005hq}.
The semi-classical algorithm was implemented as a
{\sc Fastjet} plugin {\sc ScJet}~\cite{ScPlugin} version 1.1.0.
Jet masses were calculated by summing the 4-momenta of the cells,
assuming zero mass for each cell.

\begin{figure}[tbp]
\includegraphics[width=3.4in]{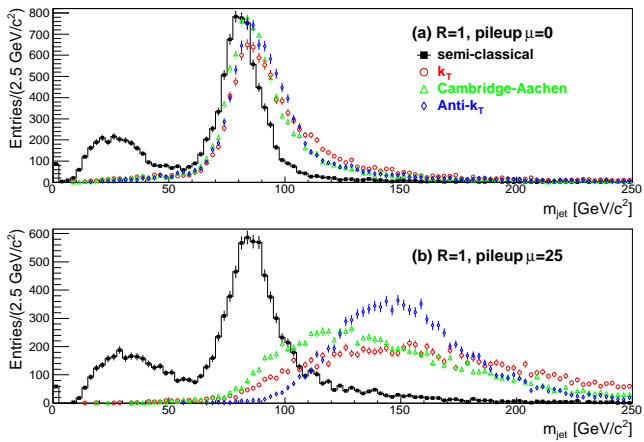}
\caption{Jet mass distributions for high-$p_T$ jets in the
same hemisphere as the generated $W$ boson, with an average of
25 pileup events overlaid and
$R=1$ for the semi-classical, $k_T$, Cambridge-Aachen, and anti-$k_T$
algorithms:  (a) no pileup; (b) with an average of 25 pileup events
overlaid.}
\label{fig:wcompare}
\end{figure}

Figure~\ref{fig:wcompare}(a) shows the jet mass distribution for jets
with $p_T>400\;{\rm GeV}/c$ in the same hemisphere as the generated $W$
for different ungroomed jet algorithms with $R=1$.  Even with no pileup,
the effect of additional radiation can be seen in the other
algorithms, while the semi-classical peak is narrowest and
lies closest, at $80.9\pm 0.1\;{\rm GeV}/c^2$,
to the generated $W$ mass of $80.385\;{\rm GeV}/c^2$.
The low and zero-mass bumps are the result of the semi-classical
algorithm ``pruning'' close
but energetically unbalanced $W$ daughters,
as noted above; combining the jet with another nearby jet
recovers the $W$ mass.  When the pileup level increases to an average of 25,
as shown in Figure~\ref{fig:wcompare}(b),
the semi-classical peak shifts roughly $4\;{\rm GeV}/c^2$ higher, but remains a
recognizable, narrow peak, while the others are much broader due
to incorporating pileup radiation.

\begin{figure}[tbp]
\includegraphics[width=3.4in]{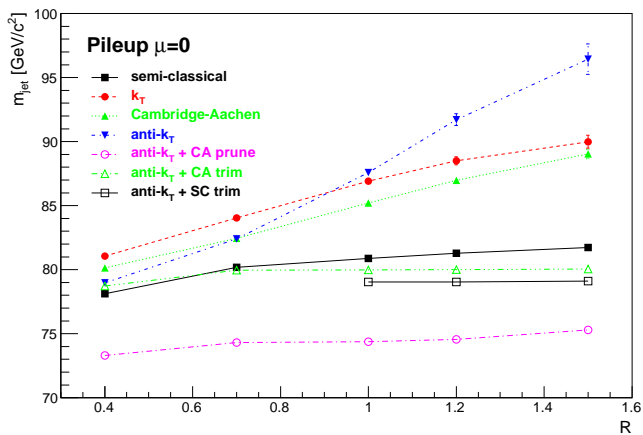}
\caption{Peak mass vs $R$ for events with zero pileup.}
\label{fig:wrcomp0}
\end{figure}

The effects of additional radiation usually are mitigated
by reducing the $R$ parameter, and indeed one can see in
Figure~\ref{fig:wrcomp0} that at $R=0.4$, all the 
peak masses cluster around $80\;{\rm GeV}/c^2$, rising
rapidly for the other ungroomed algorithms.
The semi-classical algorithm,
on the other hand, starts low at $R=0.4$,
where the two $W$ daughters often are resolved into
different jets, and levels off above $R=0.7$.

\begin{figure}[tbp]
\includegraphics[width=3.4in]{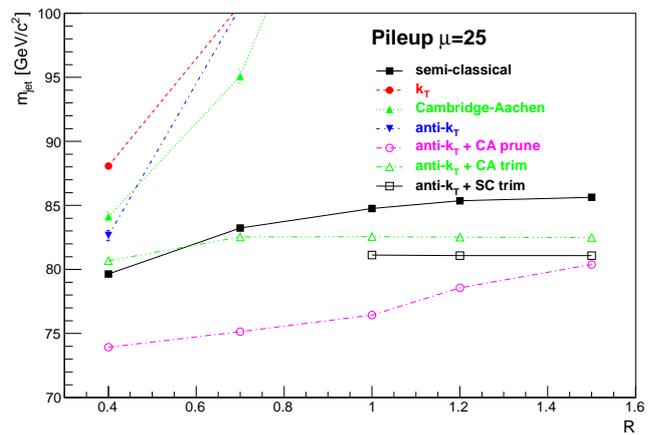}
\caption{Peak mass vs $R$ for events with average 25 pileup.
For most values of $R$ for the $k_T$, Cambridge-Aachen,
and anti-$k_T$ algorithms, the distributions are broad
rather than peaked.}
\label{fig:wrcomp25}
\end{figure}

Boosted object analyses, however, typically use large $R$ values
between 1 and $1.5$ in order to remain sensitive to a larger range of energies.
In order to mitigate pileup effects in such large jets,
the jet can undergo further ``grooming''.
It is therefore instructive to compare the new algorithm with grooming
techniques, several of which, including pruning, are also shown in
Figures~\ref{fig:wrcomp0} and \ref{fig:wrcomp25}.  It should be noted that
grooming techniques usually are tailored to particular
environments, and rely on knowledge of the target final state
such as one might use to design a search strategy based on
individually resolved jets.  The comparisons shown in this
article are therefore indicative, leaving optimization for
specific signals and backgrounds for those particular analyses.

\begin{figure}[tbp]
\includegraphics[width=3.4in]{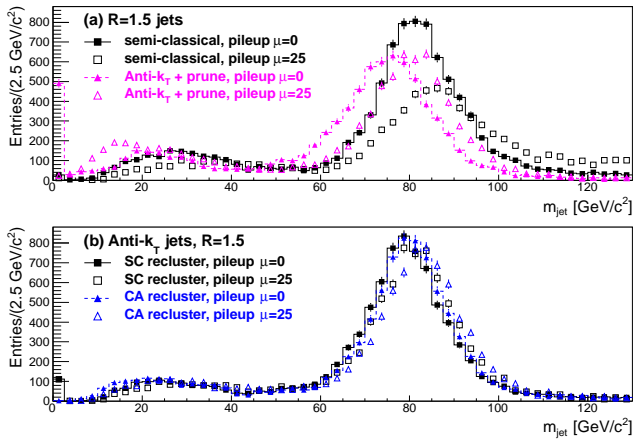}
\caption{(a) Jet mass distributions for high-$p_T$ semi-classical and
pruned anti-$k_T$ jets in the
same hemisphere as the generated $W$ boson, with zero
and average 25 pileup.
(b) Trimmed jet mass distributions, reclustered by
the Cambridge-Aachen ($R_{sub}=0.3$) and semi-classical
($R_{sub}=0.4$) algorithms, with zero and average 25 pileup.}
\label{fig:mjet_groom}
\end{figure}

Pruning has already been described.  We start with anti-$k_T$ jets
with a given $R$, and use the parameters
$z_{cut}=0.1$ and $D_{cut}=0.2$~\cite{ATLAS:2012jla} to prune.
We compare the resulting jets with those from the semi-classical
algorithm by itself, with the same $R$.
Not surprisingly, the two algorithms
behave similarly in Figures~\ref{fig:wrcomp0} and \ref{fig:wrcomp25},
even rising at
a similar rate as the average pileup level increases to 25.
Jet mass distributions for $R=1.5$
are shown in Figure~\ref{fig:mjet_groom}(a).
The presence of pileup shifts the peaks of the distributions upward,
as expected.  The semi-classical algorithm, however,
leaves a larger high-mass
tail, but also a smaller low-mass bump, suggesting that while it
eliminates less pileup radiation, it retains the $W$ daughters more often.
It is also evident that the given pruning parameters are
too aggressive for these particular conditions, resulting in a low peak mass.

Next, we consider the grooming technique of trimming,
which attempts to discern narrow, high-$p_T$ subjets within
the parent jet~\cite{Krohn:2009th,Abdesselam:2010pt}.
For the comparison, we use the Cambridge-Aachen
algorithm to recluster within the parent jet with a smaller radius
parameter $R_{sub}=0.3$, and discard the resulting subjets with
$p_T<f_{sub}P_T$, where $f_{sub}=0.05$ is a parameter and
$P_T$ is the transverse momentum of the parent
jet~\cite{ATLAS:2012jla}.
The jet mass is then calculated by summing the remaining
high-$p_T$ subjets.  Figure~\ref{fig:wrcomp25} shows
trimming to be more stable under these pileup conditions than
pruning or the ungroomed semi-classical algorithm.

The semi-classical algorithm can be used for reclustering;
in effect, such a method combines pruning and
conventional trimming.  We use a slightly larger value
of $R_{sub}=0.4$ to compensate for the smaller semi-classical
jets, and we trim anti-$k_T$ jets with $R\geq 1$.
Figure~\ref{fig:mjet_groom}(b) shows the
results of reclustering with the Cambridge-Aachen and
semi-classical algorithms.  As may be expected, the
mass distributions are very similar, again with low-mass
bumps where another $W$ daughter has been discarded
by the trimming technique.  The distributions are largely
insensitive to both the parent jet's $R$ parameter,
as also shown in
Figures~\ref{fig:wrcomp0} and \ref{fig:wrcomp25},
as well as to the pileup level.

\begin{figure}[tbp]
\includegraphics[width=3.4in]{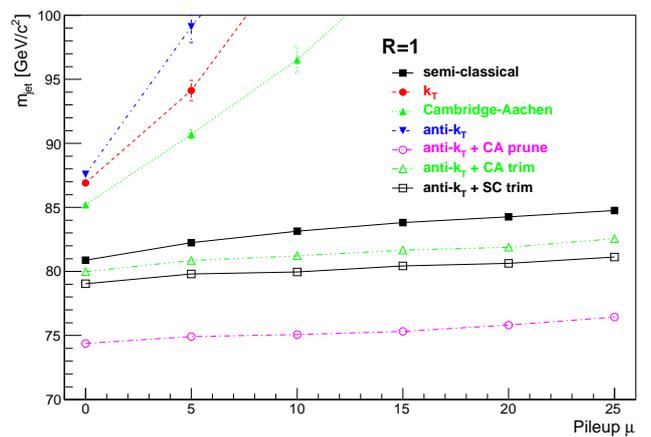}
\caption{Dependence of mass peak position on pileup for
different algorithms with $R=1$.
The mass distributions at
most or all pileup levels for the $k_T$, Cambridge-Aachen, and
anti-$k_T$ algorithms are very broad, with maxima
above $100\;{\rm GeV}/c^2$.}
\label{fig:mjet_mu_r10}
\end{figure}

\begin{figure}[tbp]
\includegraphics[width=3.4in]{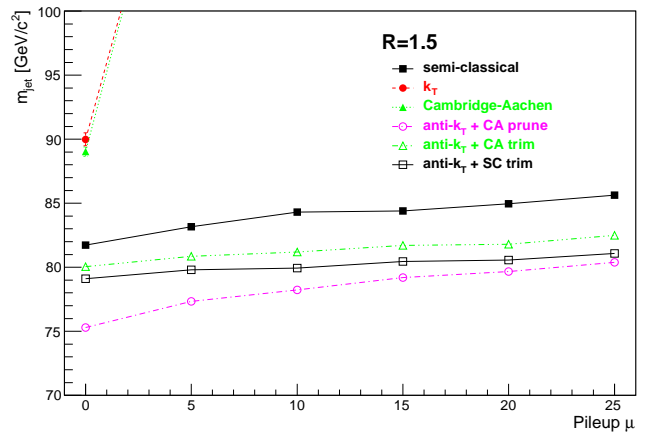}
\caption{Dependence of mass peak position on pileup for
different algorithms with $R=1.5$.
The mass distributions at
most or all pileup levels for the $k_T$, Cambridge-Aachen, and
anti-$k_T$ algorithms are very broad, with maxima
above $100\;{\rm GeV}/c^2$.  The
anti-$k_T$ mass distribution peaks near $100\;{\rm GeV}/c^2$
even at zero pileup.}
\label{fig:mjet_mu_r15}
\end{figure}

Figures~\ref{fig:mjet_mu_r10} and \ref{fig:mjet_mu_r15}
compare the effects of increasing
the pileup level on the different ungroomed and groomed
algorithms for large $R$ values.  The difference between
ungroomed and groomed jets is more obvious here,
with the mass peak rapidly rising and broadening at
even modest levels of pileup for all the ungroomed algorithms
except the semi-classical algorithm.
The ungroomed semi-classical algorithm parallels pruning over
this range of pileup level, while the peak $W$ masses of
trimming, with either reclustering algorithm,
rises more slowly than those for pruning.

\section{Conclusion}

In this article, we have investigated the behavior of a 
sequential recombination algorithm with a new
inter-cluster distance measure $d_{ij}$ which depends
on the sum of clusters' transverse masses.
The resulting algorithm effectively combines
jet clustering with pruning-like behavior in one step.
Monte Carlo tests with {\sc Pythia8} show the algorithm by itself
performing like an algorithm with jet grooming in terms of stability with
respect to the jet scale parameter $R$ as well as to pileup.
It can also be used to recluster narrow subjets for trimming.
Further work would be needed to determine whether cross sections
can be calculated for the new algorithm without large QCD corrections.
At the same time, as
has been observed widely (and wisely), Monte Carlo studies may show
the feasibility of a method, but they are a far cry from optimizing
and testing it in a genuine experimental context.

\begin{acknowledgments}
This work was supported by the Science and Technology Facilities Council
of the United Kingdom and the Higher Education Funding Council of
England.  The authors would like to thank A~Cooper-Sarkar,
C~Issever, BT~Huffman, and G~Salam for useful comments and discussion.
\end{acknowledgments}

\bibliography{r3rev}

\end{document}